\documentclass[10pt,conference]{IEEEtran}
\usepackage{courier}
\usepackage[ruled,vlined]{algorithm2e}
\usepackage{graphicx}
\usepackage{paralist}
\usepackage{tabularx}
\usepackage{balance}
\usepackage{multirow}
\usepackage{multicol}
\usepackage{pbox}
\usepackage{mathtools}
\usepackage[TABBOTCAP]{subfigure}
\usepackage{algorithmic}
\usepackage{paralist}
\usepackage{tabularx}
\usepackage{url}
\usepackage{balance}
\usepackage{multirow}
\usepackage{multicol}
\usepackage{amsmath}
\usepackage{setspace}
\usepackage{enumitem}
\usepackage[firstpage]{draftwatermark}
\usepackage{verbatim}
\usepackage{float}
\usepackage[normalem]{ulem}
\usepackage{xspace}
\usepackage{amssymb}
\usepackage{ifthen}
\usepackage{pifont}
\usepackage{tikz}
\newcommand*\circled[1]{\tikz[baseline=(char.base)]{
            \node[shape=circle,draw,inner sep=0.5pt] (char) {#1};}}
\usepackage{hyperref}
\usepackage[most]{tcolorbox}
\usepackage{booktabs}

\usepackage[normalem]{ulem}
\usepackage{colortbl}

\definecolor{MidnightBlue}{HTML}{01693F}
\include{macro}

\newcommand{\ContextDP}{{\sc ContextDP}\xspace}

\newcommand{\AidUI}{{\sc AidUI}\xspace}
\newcommand{\AidUIs}{{\sc AidUI's~}\xspace}

\newcommand{\ie}{\textit{i.e.,}\xspace}
\newcommand{\eg}{\textit{e.g.,}\xspace}

\newcommand{\etal}{\textit{et al.}\xspace}

\newcolumntype{C}[1]{>{\centering\arraybackslash}p{#1}}

\SetWatermarkAngle{0}
\SetWatermarkText{\raisebox{26cm}{%
  \hspace{16cm}%
  \href{https://www.acm.org/publications/policies/artifact-review-and-badging-current}{\includegraphics[width=0.17\linewidth]{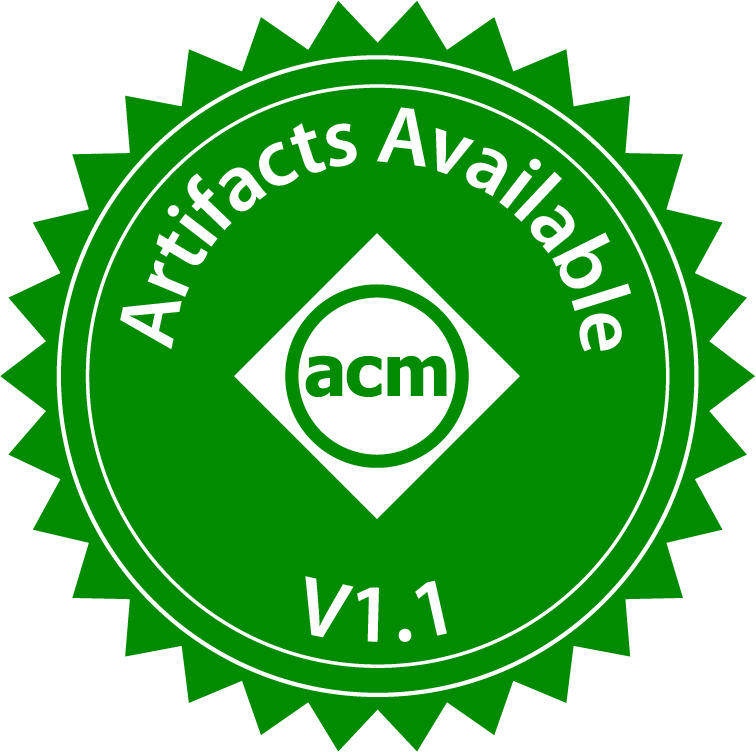}}% 
  \href{https://www.acm.org/publications/policies/artifact-review-and-badging-current}{\includegraphics[width=0.17\linewidth]{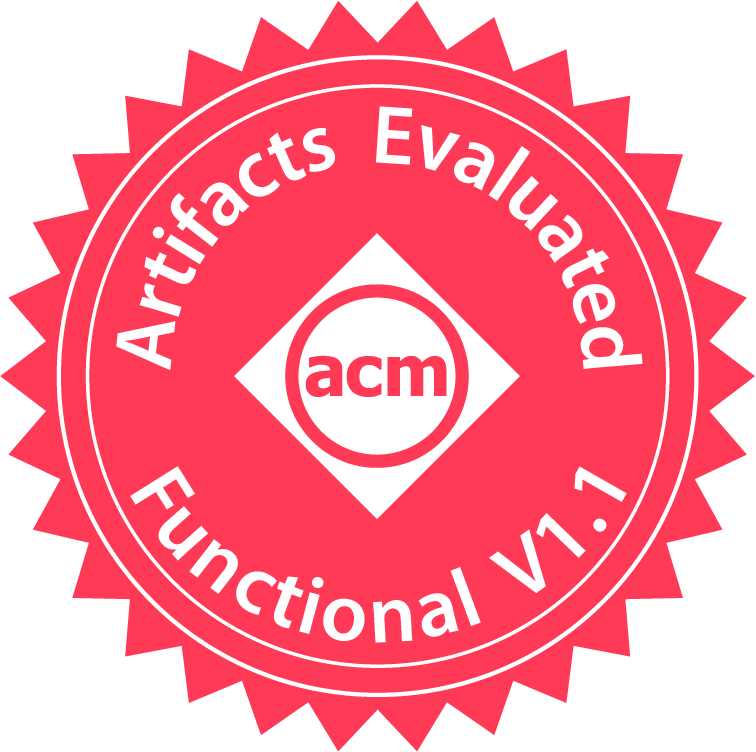}}%
}}

\begin{document}
\title{
AidUI: Toward Automated Recognition of \\Dark Patterns in User Interfaces
}


\author{\IEEEauthorblockN{S M Hasan Mansur, Sabiha Salma}
\IEEEauthorblockA{Department of Computer Science\\
George Mason University\\
Fairfax, VA\\
Email: \{smansur4,ssalma\}@gmu.edu}
\and
\IEEEauthorblockN{Damilola Awofisayo\IEEEauthorrefmark{2}}
\IEEEauthorblockA{Department of Computer Science\\
Duke University\\
Durham, NC\\
Email: dami.awofisayo@duke.edu}
\and
\IEEEauthorblockN{Kevin Moran}
\IEEEauthorblockA{Department of Computer Science\\
George Mason University\\
Fairfax, VA\\
Email: kpmoran@gmu.edu}
}

\maketitle
\begingroup\renewcommand\thefootnote{\IEEEauthorrefmark{2}}
\footnotetext{Work completed as a high school student at George Mason University's Aspiring Scientist Summer Internship Program (ASSIP).}

\begingroup\renewcommand\thefootnote{1}

%Adding this temporarily to show page numbers
\thispagestyle{plain}
\pagestyle{plain}

\begin{abstract}

Past studies have illustrated the prevalence of \textit{UI dark patterns}, or user interfaces that can lead end-users toward (unknowingly) taking actions that they may not have intended. Such deceptive UI designs can be either intentional (to benefit an online service) or unintentional (through complicit design practices) and can result in adverse effects on end users, such as oversharing personal information or financial loss. While significant research progress has been made toward the development of dark pattern taxonomies across different software domains, developers and users currently lack guidance to help recognize, avoid, and navigate these often subtle design motifs. However, automated recognition of dark patterns is a challenging task, as the instantiation of a single type of pattern can take many forms, leading to significant variability.

In this paper, we take the first step toward understanding the extent to which common UI dark patterns can be \textit{automatically} recognized in modern software applications. To do this, we introduce \AidUI, a novel automated approach that uses computer vision and natural language processing techniques to recognize a set of visual and textual \textit{cues} in application screenshots that signify the presence of ten unique UI dark patterns, allowing for their detection, classification, and localization. To evaluate our approach, we have constructed \ContextDP, the current largest dataset of fully-localized UI dark patterns that spans 175 mobile and 83 web UI screenshots containing 301 dark pattern instances. The results of our evaluation illustrate that \AidUI achieves an overall precision of 0.66, recall of 0.67, F1-score of 0.65 in detecting dark pattern instances, reports few false positives, and is able to localize detected patterns with an IoU score of ~0.84. Furthermore, a significant subset of our studied dark patterns can be detected quite reliably (F1 score of over 0.82), and future research directions may allow for improved detection of additional patterns. This work demonstrates the plausibility of developing tools to aid developers in recognizing and appropriately rectifying deceptive UI patterns.

\end{abstract}


\section{Introduction}
\label{introduction}

Modern user interfaces (UIs) have unprecedented influence on the daily lives of users due to the increasing digitization of common tasks -- ranging from financial transactions to shopping. As such, an emphasis on ease of use in the design of these interfaces has never been more critical. However, while UI designers typically strive to create interfaces that facilitate seamless completion of computing tasks, they are also capable of creating interfaces that may subtly mislead users into performing tasks they did not intend, but which may benefit business stakeholders. This dichotomy is influenced by competing design pressures: (i) designing easy to use UIs may increase the reputation and overall market share of an application; and (ii) designing UIs that intentionally influence users into performing certain actions may benefit the goals of application stakeholders.

\begin{figure}
\centering
\vspace{1em}
\includegraphics[width=0.80\linewidth]{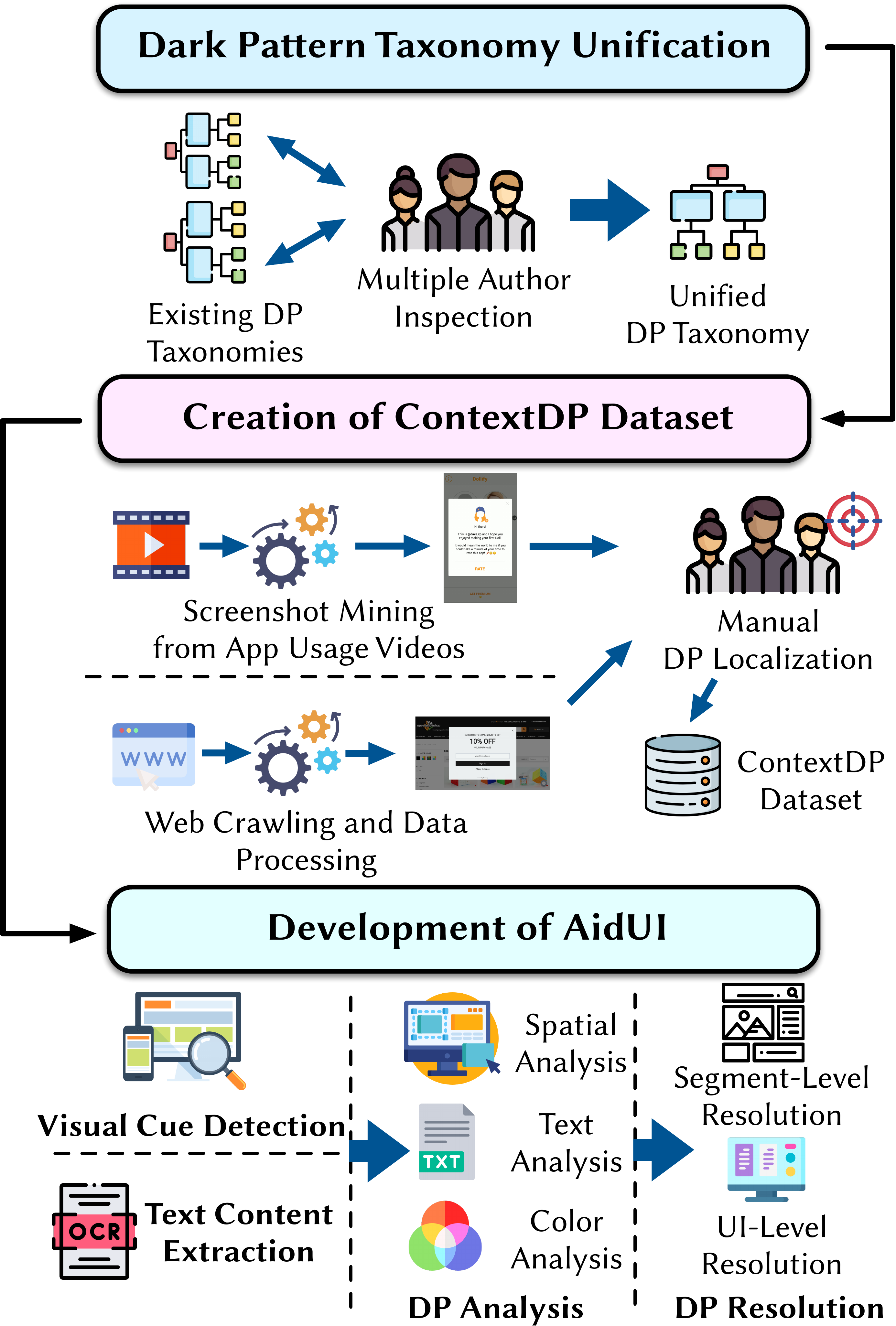}
\caption{This paper presents (i) a unified taxonomy of UI dark patterns, (ii) the \ContextDP dataset which contains 501 screenshots depicting 301 DP and 243 non-DP instances, and (iii) the \AidUI approach, which is able to identify and process visual and textual cues to detect and localize dark patterns.}
\vspace{1em}
\label{fig:overview}
\end{figure}

This deceptive side of UI design has received increasing attention from various research communities in recent years, and has led to an increased shared understanding of a phenomenon referred to as \emph{dark patterns (DPs)}. While the notion of DPs appeared in research literature as early as 2010~\cite{brignull}, Mathur ~\etal~\cite{mathur2019dark} recently defined dark patterns as:

\vspace{0.5em}
 \noindent\textit{``User interface design choices that benefit an online service by coercing, steering, or deceiving users into making decisions that, if fully informed and capable of selecting alternatives, they might not make''}.
\vspace{0.5em}

\noindent To further illustrate the concept of a dark pattern, we provide an illustrative example of the \texttt{\small Attention Distraction} pattern in Figure~\ref{fig:attention-distraction-example} where the size and color of a UI element is used to draw the user's attention and influence them into making a particular choice. In our example, as illustrated in Figure~\ref{fig:attention-distraction-example}, there is a significant size difference between the text ``Continue'' \& ``skip seat selection'' options, making it more likely that a user might pay for a seat as opposed to forgoing the selection and being automatically assigned one for free.

\begin{figure}
\centering
%\vspace{1em}
\includegraphics[width=\linewidth]{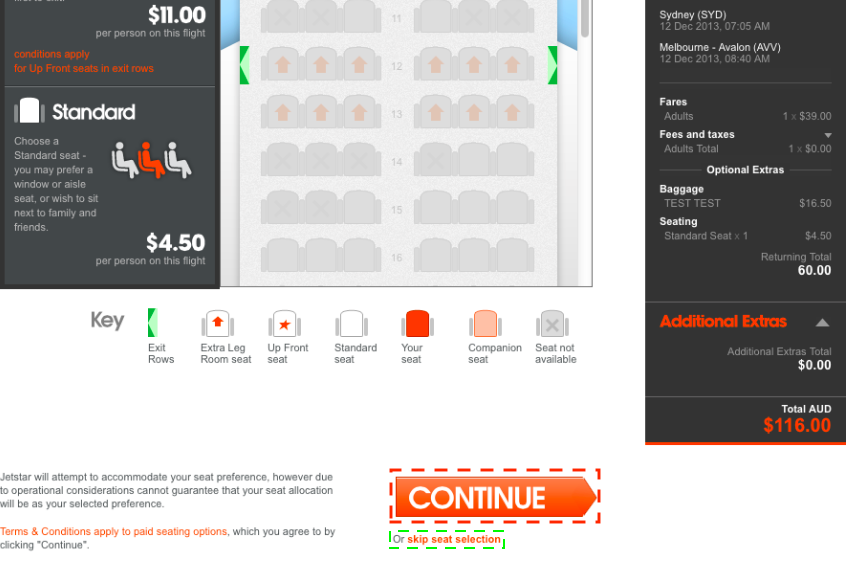}
%\vspace{-1em}
\caption{Example of an \texttt{\scriptsize Attention Distraction} Dark Pattern}
\label{fig:attention-distraction-example}
\end{figure}

There is ongoing work from the human-computer interaction (HCI) community into understanding the ethical considerations~\cite{gray2018dark,gray2020assholes}, constructions~\cite{greenberg2014dark,zagal2013dark,lewis2014patterns,bosch2016tales,lacey2019cuteness,utz2019informed,westin2019opt-out,gray2020assholes}, user perspectives~\cite{conti2010malicious,gray2020assholes,maier2020dark}, and practitioner perspectives~\cite{gray2018dark} of DPs. Additionally, the community has worked to construct evidence-based taxonomies of generally agreed upon DPs present in modern UIs~\cite{greenberg2014dark,zagal2013dark,lewis2014patterns,bosch2016tales,lacey2019cuteness,utz2019informed,westin2019opt-out,gray2020assholes,gray2018dark,mathur2019dark}. The prevalence of these identified categories of DPs is undeniable, as recent studies have illustrated that nearly 11\% of top shopping websites~\cite{mathur2019dark} and nearly 95\% of the top 240 apps on Google Play~\cite{di2020ui} contain one or more defined DPs. It is important to note that such DPs may not always be the result of ill intent from a designers vantage point, and past work has suggested that the current prevalence of such patterns may be the result of complicit, and perhaps unintentional, design practices on the part of software designers and developers~\cite{gray2018dark}.

However, regardless of the reason for their introduction and continued use, it is clear that deceptive UI patterns can have a negative effect on end-users. Such effects have been documented by prior work illustrating that, in many cases, DPs can result in financial loss~\cite{johnson_2013, asbury_2014}, or in oversharing personal information~\cite{fansher2018darkpatterns,zagal2013dark}. Additionally, a prior human study that examined end-user perception of DPs in mobile apps found that most users cannot recognize DPs~\cite{di2020ui}. Given the prevalence and potential negative effects of deceptive UI patterns, and past evidence that designers and developers may unintentionally introduce such designs into their apps~\cite{gray2018dark}, developer-facing tools that can automatically recognize and signal the presence of such patterns could aid in avoiding these deceptive designs and improve overall software quality.

\begin{figure*}[t]
\centering
\vspace{-1em}
\includegraphics[width=\linewidth]{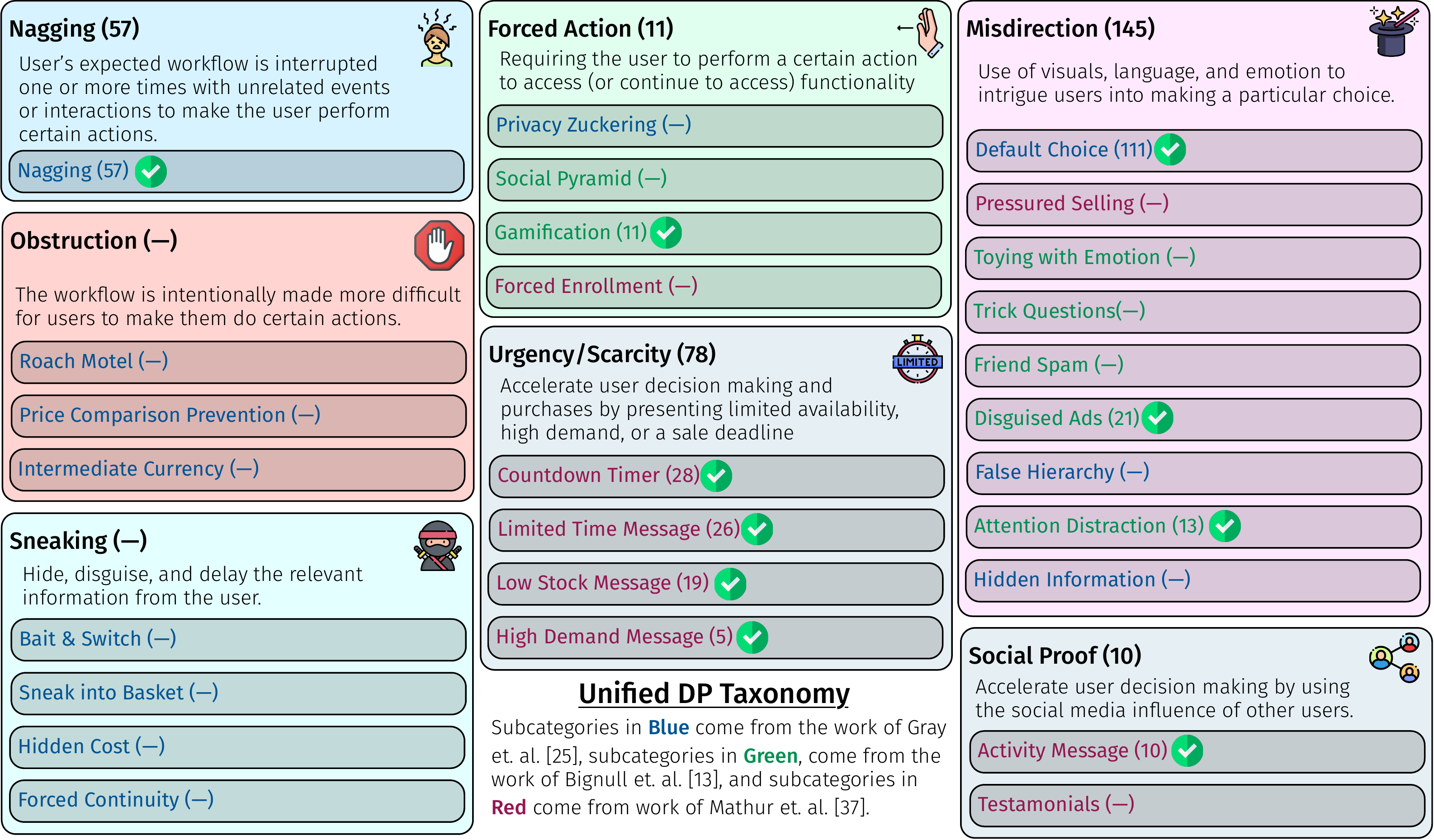}
\vspace{-0.5em}
\caption{Our Unified Dark Pattern Taxonomy -- numbers in parentheses signify the number of examples of the DPs that are present in the \ContextDP dataset, and the \img{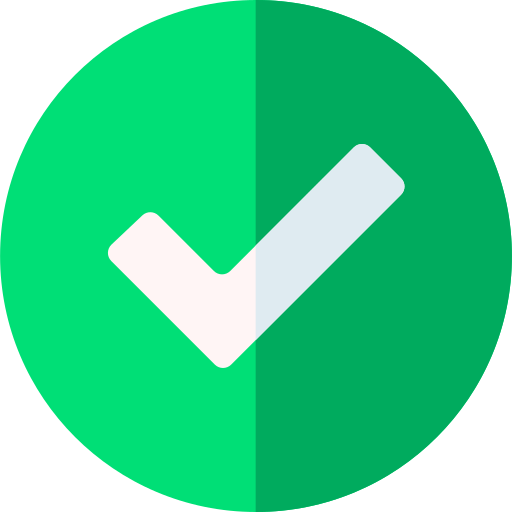} signifies those patterns that \AidUI is designed to detect. Note that this unified taxonomy is a targeted combination of past works~\cite{mathur2019dark,gray2018dark,brignull} with minor modifications. }
\vspace{-1em}
\label{fig:taxonomy}
\end{figure*}

Despite the potential benefit of automated techniques for detecting DPs, there are two key challenges related to the design and implementation of such an approach. First, while the research community has made considerable progress in defining various types of DPs across a number of domains~\cite{di2020ui,mathur2019dark,brignull,gray2018dark}, there can still be significant variability in how these patterns are \textit{instantiated} in software applications. That is, while the very etymology of the phrase ``dark pattern'' suggests the presence of strong semantic signals that characterize different deceptive UI designs, the actual implementation of such designs can vary significantly. This makes designing an approach to detect such patterns difficult. Second, the research community currently lacks a large dataset of DPs with \textit{fine-grained localization information} mapped to a \textit{unified} taxonomy of DPs. While prior work has led to the creation of existing datasets, such as the one produced by Mathur~\etal~\cite{mathur2019dark} related to online shopping, these datasets do not contain localized DPs (\eg bounding boxes that denote the location and spatial properties of the DPs) and such datasets are typically created with domain-specific categories of DPs.

To address the need for the automation of the detection of DPs in UIs, we introduce \AidUI (\textbf{Aid} for detecting \textbf{UI} Dark Patterns) - an approach that conducts \textit{textual}, \textit{icon}, \textit{color}, and \textit{spatial} analysis of a UI to automatically detect the presence of underlying DPs. The key idea underlying AidUI is that there exist several visual and textual \textit{cues}, that when (co)-appearing, signify the presence of various dark patterns. By detecting these individual cues, and analyzing their (co)-occurrence, \AidUI is able to overcome challenges related to the variability of dark patterns as they appear ``in-the-wild''. \AidUI is the first approach to attempt to detect DPs using a \textit{fully automated} process, and performs cue detection using a combination of computer vision and natural language template matching techniques. 
\AidUI operates solely on \textit{visual} data, requiring \textit{only a screenshot of a user interface as input}, making it easily extensible to multiple software domains.

In the process of building and evaluating \AidUI we have constructed \ContextDP, the current largest dataset of fully-localized DP instances to UI screenshots, containing 258 screenshots with 301 DP instances. The DP instances map to a \textit{unified set of ten DP categories} derived through merging common DP categories described in previous taxonomies~\cite{mathur2019dark,gray2018dark,brignull}. This dataset spans multiple application domains, including 175 mobile app screenshots and 83 Web UI screenshots, for a total of 258 screenshots depicting 301 DPs. Additionally, we augment our dataset with a set of 243 screenshots (164 mobile \& 79 web) that do not contain DPs to investigate the likelihood of AidUI to detect false positives.

We conducted a comprehensive evaluation to measure the effectiveness and robustness of \AidUI. Using \ContextDP, we measured the precision, recall and F1-score of \AidUI's DP detection capabilities, as well as its localization performance in different settings. Additionally, we performed an ablation study to examine which components of our technique contributed most to the overall detection performance. We found that, across all DP instances in our dataset, \AidUI achieved an overall average precision of 0.66, average recall of 0.67 and average F1-score of 0.65. However, for five DPs, \AidUI was able to achieve higher precision (0.87), recall (0.80) and F1-score (0.82) compared to the five other DP categories, illustrating the variance in difficulty of detecting different types of patterns. Our analysis of localization performance illustrates that AidUI can provide useful localization that is specific to the various salient patterns it detects. Finally, our ablation study illustrated that combining text analysis with the color and spatial analysis led to the highest performance. 

\noindent In summary, the contributions of our work are as follows:
\begin{itemize}
    \item \AidUI, the first automated approach capable of detecting the presence of a diverse set of DPs. To accomplish this, the proposed approach adapts techniques from natural language processing and computer vision to detect various cues that signify dark patterns.
    \item \ContextDP, which is the current largest dataset of DP instances localized to UI screenshots. \ContextDP contains 162 web and 339 mobile screenshots depicting 301 DP and 243 Non-DP instances.
    \item A comprehensive empirical evaluation of \AidUI that measures the precision, recall, F1-score, localization performance, and the robustness of the approach. Our evaluation illustrates that \AidUI can detect DPs from a subset of our studied DP categories with higher precision, recall and F1-score, with useful localization results.
    \item Online appendices~\cite{appendix,zenodo,github} containing source code, experimental data, and trained models that can facilitate the replication of our results and encourage future work on automated detection of UI DPs.
\end{itemize}

%\vspace{-1em}
\section{Unifying Web and Mobile Dark Patterns}
\label{background}

\begin{figure*}
\centering
\vspace{-1em}
\includegraphics[width=0.9\linewidth]{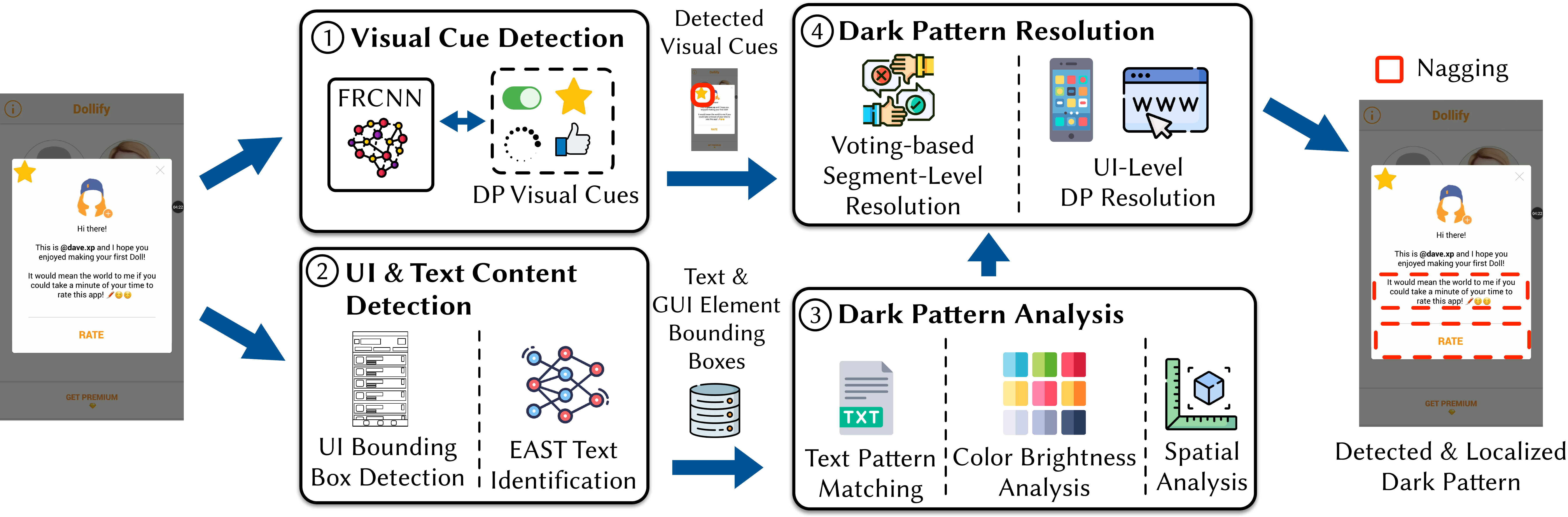}
\vspace{-1em}
\caption{Workflow of the \AidUI approach}
\vspace{-1em}
\label{fig:aidui_approach}
\end{figure*}

There has been a wealth of work from the general HCI community that has constructed DP taxonomies. One of the earliest taxonomies comes from Brignull~\cite{brignull} who not only coined the term ``Dark Pattern'', but also proposed a classification of DPs into different categories. The DP categories originally derived by Brignull~\etal~\cite{brignull} are available on the \texttt{\small darkpatterns.org}\footnote{Note that since the publishing of this paper, Brignull moved to using the term ``deceptive design'' in place of ``dark pattern'' in an effort to be clearer and more inclusive.} portal with examples from web and mobile applications. Gray~\etal~\cite{gray2018dark} redefined Brignull's taxonomy and proposed a new set of patterns consisting of five main categories of DPs that include (i) Nagging, (ii) Obstruction, (iii) Sneaking, (iv) Interface Interference and (v) Forced Action. Geronimo~\etal~\cite{di2020ui} used Gray's taxonomy and extended the original meaning of Aesthetic Manipulation and Forced Action classes to include new DPs instances. The authors analyzed pervasiveness of DPs in mobile applications using an approach similar to cognitive walk through techniques~\cite{nielsen1994usability} whereas previous works \cite{gray2018dark, mathur2019dark, moser2019impulse} analyzed screenshots to classify DPs. Finally, Mathur~\etal \cite{mathur2019dark} performed a semi-automated, large-scale collection of DPs in online shopping websites, and derived a taxonomy of 15 dark patterns grouped into 7 categories. In the course of their data collection, they found 1,818 instances of DPs on the top $\approx$11k shopping websites. More recent work has aimed to identify DPs that are specific to various different contexts or domains including (i) shopping web apps~\cite{mathur2019dark}, (ii) computer games~\cite{zagal2013dark}, (iii) privacy-centric software~\cite{bosch2016tales}, (iv) robotics~\cite{lacey2019cuteness}, and (v) digital consent forms~\cite{utz2019informed}.

Given the somewhat complementary, yet disparate nature of existing taxonomies of DPs, we aimed to merge similar DP categories from existing taxonomies together and provide  a larger landscape of patterns for mobile and web apps toward which we can design and evaluate our automated detection approach. Given that the scope of our work is primarily concerned with web and mobile UIs, we did not include many of the domain specific taxonomies mentioned above. Instead, our unified taxonomy is primarily a fusing of the various categories and subcategories derived by Gray~\etal~\cite{gray2018dark}, Mathur~\etal~\cite{mathur2019dark} and Brignull~\etal~\cite{brignull}.
To build our unified taxonomy, first one author gathered one-two existing examples of each type of DPs that exists in the categories described by Gray~\etal~\cite{gray2018dark}, Mathur~\etal~\cite{mathur2019dark}, and Birgnull~\etal~\cite{brignull}. Next, two authors met to review each DP example and re-group the various categories under unified headings. Our final unified taxonomy, illustrated in Figure~\ref{fig:taxonomy}, spans 7 parent categories which organize a total of 27 classes that describe different DPs. Note that many of the DPs in our taxonomy are self-documenting (\eg \texttt{\small countdown timer}), however, we provide full descriptions and examples of each DP in our online appendices~\cite{appendix,github,zenodo}.

Not all dark patterns are created equal from viewpoint of the underlying UI motifs that signal their presence. For example, for the \texttt{\small Sneak Into Basket} DP type, typically there would be several actions and screens required to detect the presence of such a pattern, which introduces far more variability in its potential observed visual and textual cues. With this in mind we aimed to prioritize the detection strategy of \AidUI toward certain patterns that carry with them distinct visual and textual cues which both manifest on a \textit{single screen}.
We leave the detection of dynamic DPs involving multiple screens and actions for future work. To perform this prioritization, two authors met and further discussed the examples of each of the 27 classes of DPs and marked those that exhibited salient visual and textual cues, noting these cues for use in the later implementation of \AidUI. Thus, we identified a final set of 10 target DPs , toward which we oriented \AidUIs analysis. The targeted DP categories are marked with a \img{images/check.png} in Figure~\ref{fig:taxonomy}.

\section{The A{\small ID}UI Approach}

This section presents the \AidUI approach for automatically detecting DPs in UIs. The architecture of \AidUI, depicted in Figure \ref{fig:aidui_approach}, is designed around four main phases: \circled{1} the \emph{Visual Cue Detection} phase, which leverages a deep learning based object detection model to identify UI objects representing visual cues for DPs, \circled{2} the \emph{UI \& Text Content Detection} phase, which extracts UI segments containing both text  and non-text content, \circled{3} the \emph{DP Analysis} phase, which employs text pattern matching, as well as color and spatial analysis techniques to analyze the extracted UI segments and identify a set of potential DPs,  and \circled{4} the \emph{DP Resolution} phase, which uses results from both Visual Cue Detection and DP Analysis phases to predict a final set of underlying DPs in a given UI. It is important to note that \AidUI operates \textit{purely on pixel-based data} from UI screenshots, making it extensible to different software domains. In the remainder of this section, we first discuss the motivation of the architecture of our approach and then discuss each of the four phases in detail.

\subsection{Approach Motivation}
\label{sec: Approach Background}

While studying existing DP taxonomies, we observed that different DP categories tend to include certain types of icons and text as well as exhibit distinct patterns in color brightness and spatial organization of UI elements. For example, instances from \texttt{\small nagging} category are likely to have both visual (\ie rating related icons such as like buttons, stars, etc.) and textual cues (\ie keywords such as "rate", "rating" etc.). Based on these observations regarding the visual and textual cues across different DP categories, we have identified five tasks that are required for detecting DPs. These include two detection tasks (\ie visual cue detection and text content detection) as well as three analysis tasks related to properties of UI components (\ie text, color, and spatial analysis). The major phases of our approach are designed around these detection and analysis tasks.

\subsection{Phase 1: Visual Cue Detection}
\label{sec: Visual Cue Detection}
The main goal of this phase is to identify the icons that can serve as important visual cues for detecting DPs. To accomplish this goal, our approach leverages deep learning based object detection techniques. In this case, we adapt an implementation of Faster R-CNN \cite{ren2015faster} to accurately identify and localize specific types of icons in a target UI.

The Visual Cue Detection phase receives a UI as an input, then uses Faster R-CNN model to detect the positions and bounding boxes of the target icons. Finally, through a mapping of detected icons to likely candidate DPs, it provides a list of potential DP categories. The output of this phase is a \texttt{\small json} file consisting of bounding boxes and DP categories for detected icons with the highest confidence scores.

\subsubsection{Icon Detection}
Typically, DL models such as the Faster R-CNN model are data hungry and need to be trained with manually labeled, large datasets. However, for the visual cues that we wish to detect (\eg rating bars, like buttons, etc.), we only have a small set of examples. Labeling such UI components in existing UI datasets, such as RICO~\cite{Deka:2017:Rico}, would be extremely time consuming, and hence would not scale. As such, we develop a fully-automated training data generation technique that allows for the creation of large sets of training data for our target visual cues. This training data generation technique has been successfully used in past work~\cite{cardenas-translating} to detect touch indicators on mobile UI screenshots with extremely high accuracy (\eg 98\%). It should be noted that this data generation procedure is completely independent and separate from the \ContextDP dataset. That is, we automatically generate the dataset to train our neural icon detection model such that no screens or apps used in the training of this model appear in the \ContextDP dataset.

We used the existing large scale RICO \cite{Deka:2017:Rico} UI dataset to aid in our data generation process. RICO, by far the largest repository of mobile app designs, contains 66k unique UI screens from 9.3k free Android apps spanning over 27 categories. First, we randomly sampled 1020 unique UIs of different apps from the RICO dataset. NExt, we randomly sampled different versions of the icons that we target to detect as visual cues which are freely available on icon or UI design websites. In total, we collect total 16 different versions of 6 icon types (\ie 2 Google Ad icons, 3 loading icons, 3 like icons, 3 dislike icons, 2 star icons, and 3 toggle switch icons). Next, we programmatically overlay the icons on the UIs to create a dataset consisting of total 16320 images (\ie 16 icons x 1020 UIs) and corresponding annotations in MSCOCO \cite{lin2014microsoft} format. Note that past work has illustrated a dataset of $\approx$ 16k screenshots to be sufficient to train a neural object detection approach to detect icons~\cite{cardenas-translating}. While creating synthesized data, we place randomly sized icons (\ie foreground image) at random locations on the UIs (\ie background image). To create training and validation sets, we split the dataset into 80\%/20\% respectively. We then use the open source machine learning framework PyTorch \cite{pytorch} and Torchvision \cite{torchvision} library to adapt an implementation of the Faster R-CNN \cite{ren2015faster} object detection model. Our trained model achieves an overall accuracy of 91\% on our held out test set. The output of the object detection model at inference time for a given input screenshot is at least one, and potentially multiple, bounding boxes of identified icons, where the model assigns each bounding box a list of potential icon categories ranked by confidence level. AidUI uses \textit{all} identified bounding boxes for a given screenshot, and uses the category with the highest confidence score from the model to identify the icon category for each bounding box.

\begin{table}[t]
\vspace{-1em}
\caption{Mapping of Icons (visual cues) to likely DPs}
\label{tab:icon_to_dp_mapping}
\footnotesize
\centering
\begin{tabular}{@{}ll@{}}
\toprule
\textbf{Icons} & \textbf{Likely DPs} \\ \midrule
Like \begin{minipage}{.09\textwidth}
      \includegraphics[width=0.2\linewidth]{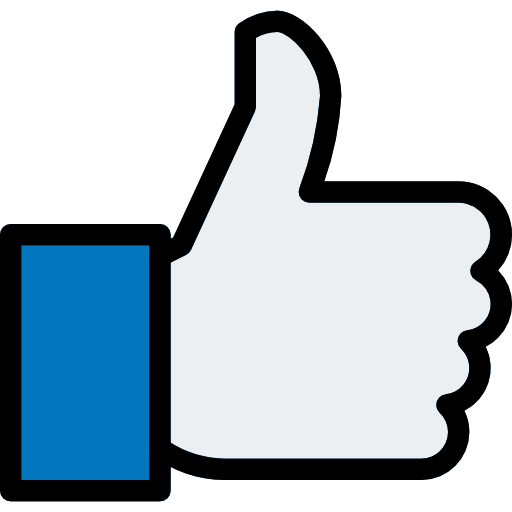}
    \end{minipage} & Nagging \\
Dislike \begin{minipage}{.09\textwidth}
      \includegraphics[width=0.2\linewidth]{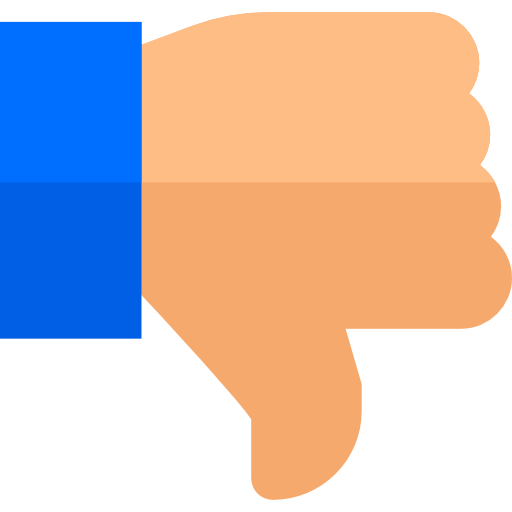}
    \end{minipage} & Nagging \\
Star \begin{minipage}{.09\textwidth}
      \includegraphics[width=0.2\linewidth]{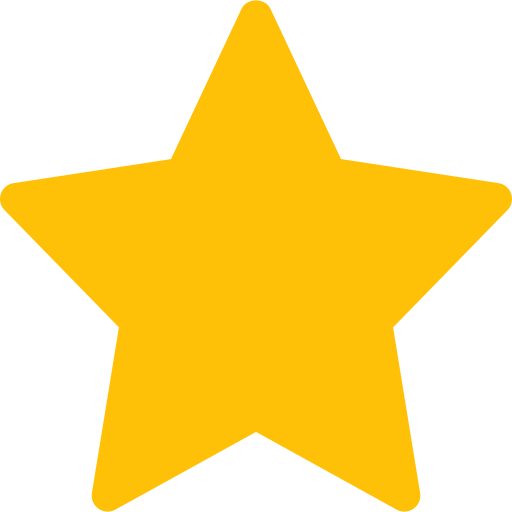}
    \end{minipage} & Nagging \\
Toggle Switch (on) \begin{minipage}{.09\textwidth}
      \includegraphics[width=0.2\linewidth]{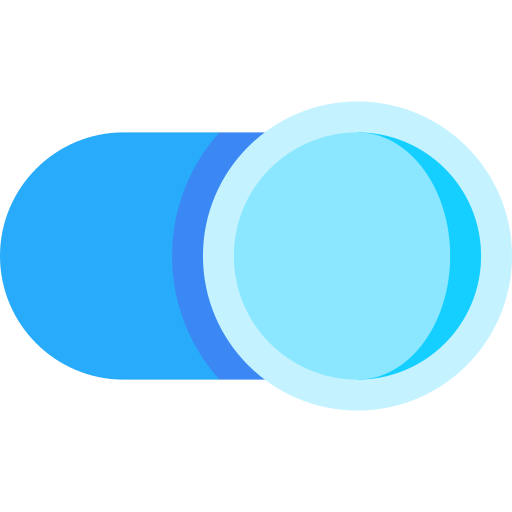}
    \end{minipage} & Default Choice \\
Ad \begin{minipage}{.09\textwidth}
      \includegraphics[width=0.2\linewidth]{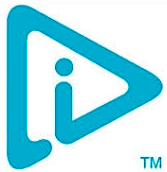}
    \end{minipage} & Nagging, Disgiused Ads \\
Ad Loader \begin{minipage}{.09\textwidth}
      \includegraphics[width=0.2\linewidth]{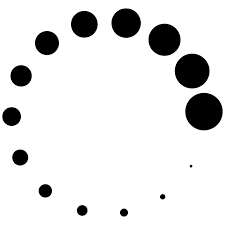}
    \end{minipage} & Nagging,  Gamification \\ \bottomrule
\end{tabular}
\vspace{.5em}
\end{table}

\subsubsection{Mapping Detected Icons to DPs}
As stated earlier in this section, some icons tend to relate to different types of DPs. Based on this observation, we translate the relationship between certain icons and DPs into a set of mapping rules (illustrated in Table~\ref{tab:icon_to_dp_mapping}). Our approach uses these predefined rules for mapping the detected icon to potential DP(s). Thus, as a final output of \textit{Visual Cue Detection} phase, a \texttt{\small JSON} file is produced that includes labels, bounding boxes and confidence scores of the detected icon(s) as well as a set of related DPs that are likely to be present in the UI. It should be noted that, \AidUI never uses \textit{only} icon detection to determine whether there is a DP, but instead combines icon detection with the other visual and textual cues(section ~\ref{sec: DP Analysis}) for DP prediction.

\subsection{Phase 2: UI \& Text Content Detection}
\label{sec: Text Content Detection}

To identify the general UI components and textual cues related to DPs, we first need to extract the UI segments that have textual content. To accomplish this, we adapted the implementation of the approach introduced by Xie~\etal~\cite{10.1145/3368089.3417940}, called UIED, that consists of two parts for detecting graphical and textual UI elements. For text detection and extraction, it leverages the EAST text detection model~\cite{zhou-east}.
For graphical UI elements, it uses an unsupervised edge detection algorithm and CNN to locate and classify elements. As we are interested in extracting UI segments with text contents, our approach collects the OCR'd output in \texttt{\small JSON} format from \textit{UIED}. The \texttt{\small JSON} file contains the text contents along with the corresponding bounding box information and serves as an input for the next phase (\ie \textit{DP Analysis}).

\subsection{Phase 3: DP Analysis}
\label{sec: DP Analysis}

In \textit{DP Analysis} phase, the main goal is to apply different analysis techniques on the UI segments containing textual information to predict the potential underlying DPs associated with those segments. As stated earlier in section \ref{sec: Approach Background}, different DP categories tend to exhibit certain textual, visual and spatial patterns. Hence, \textit{DP Analysis} phase incorporates different techniques for analyzing text, color, and spatial information extracted from the identified UI segments.

First, the UI text segments are provided as inputs. Then, text analysis is used to select the segments that have text contents showing patterns related to different DP categories. Next, color analysis is performed to categorize these selected segments based on their brightness level. Finally, in the spatial analysis, relative size information of neighbor segments is computed. The final output of the \textit{DP Analysis} phase is a \texttt{\small JSON} file that combines the results from textual, color, and spatial analysis of the UI segments which is later analyzed during DP Resolution.

\begin{table}[]
\vspace{-1em}
\centering
\scriptsize
\renewcommand{\arraystretch}{1.3}
\vspace{-1em}
\caption{Mapping of Sample Lexical Patterns to DPs}
\label{tab:text-patterns}
\begin{tabular}{p{3.6cm}p{4.5cm}}
\hline
\textbf{Dark Pattern}                 & \textbf{Sample Lexical Patterns}                                                                                                                                                                                  \\ \hline
Nagging                               & watch + \textless{}ad/session\textgreater{}                                                                                                                                                                       \\
Gamification                          & \begin{tabular}[c]{@{}l@{}}\textless{}ask/invite/refer\textgreater + friends + to + {[}subject{]}\\ signup + for + \textless{}credits/points/tokens\textgreater{}\end{tabular}                                    \\
Default Choice                        & \begin{tabular}[c]{@{}l@{}}I + \textless{}agree/consent\textgreater + to + [predicate]\end{tabular} \\
Attention Distraction                 & I +\textless{}decline/don't\textgreater + \textless{}want/opt out/refuse to\textgreater + {[}predicate{]}                                                                                                         \\
Countdown~Timer/ Limited Time Message & sale + \textless{}ends/countdown/now\textgreater{}; shop + \textless{}now/within\textgreater{}                                                                                                                    \\
Low Stock/ Limited Time/ High Demand  & \textless{}apply/order\textgreater + by; only + {[}number{]} + in stock; {[}number{]} + available                                                                                                                 \\
Activity Message                      & {[}number{]} + items sold + {[}number{]} + time                                                                                      \\ \hline
\end{tabular}
\end{table}

\subsubsection{\underline{Text Analysis}} To select the UI segments exhibiting specific textual cues, our approach leverages a pattern matching technique. Based on the observed textual cues for different DP categories, we define heuristic pattern matching rules. We defined corresponding lexical patterns for each of the targeted 10 DP categories that match keywords as well as sentence patterns. We provide a subset of these lexical patterns in Table \ref{tab:text-patterns} and provide a detailed list of these keywords and patterns in our online appendices~\cite{appendix,github,zenodo}. When a match occurs, our approach returns a \texttt{\small JSON} object containing the segment information (\ie UI coordinates, width, height etc.) of the matched text content and the corresponding DP category related to the pattern rule used. In the case that there are multiple matched contents under the same DP category, the content with the longest sequence is selected. We implement our linguistic pattern matching using spaCy \cite{spaCy}, python library.

\subsubsection{\underline{Color Analysis}} Next, in the \textit{Color Analysis} step, we categorize the detected segments from \textit{Text Analysis} step based on their brightness. To accomplish this, our approach uses a color histogram analysis technique. In this analysis process, we first calculate the grayscale histogram of the segment where we use two bins, one for the intensity values in the range of 0-127 and another one for the intensity values ranging from 128-255. If 65\% or more of the pixels have values in the range of 0-127, the segment is categorized as \textit{"darker"} segment, whereas if 65\% or more of the pixels have values in the range of 128-255, the segment is considered as a \textit{"brighter"} one. In all other cases, the segment is categorized as a \textit{"normal"} one. The reason for performing this classification of color intensity is that there are some DP types (such as \texttt{\small Attention} \texttt{\small Distraction} and \texttt{\small Default Choice}) wherein orthogonal brightness/contrast levels are likely to signal the presence of a DP. To implement color analysis, we use the Python API for OpenCV \cite{OpenCV} to perform a color histogram analysis that provides the brightness measure mentioned earlier. For each segment, the \textit{Color Analysis} step outputs a \texttt{\small JSON}  object containing the histogram results and the associated brightness category (\ie darker/brighter/normal).

\subsubsection{\underline{Spatial Analysis}} Some DP categories (such as \texttt{\small Attention} \texttt{\small Distraction} and \texttt{\small Default Choice}) are signified by size differences between their nearest components. Hence, in the \textit{Spatial Analysis} step, our main goal is to find the neighbor segments around a given segment and compute the relative size of a segment in comparison to the size of the neighbors. Here, our approach conducts a two step spatial analysis. First, for a given segment, we calculate the neighborhood area around it by adding a proximity factor to the segment boundaries. The proximity factor is calculated as a very small percentage (\eg $\approx 5\%$) of the segment size, which is derived empirically. If any other segment's boundary intersects with the boundary of the neighborhood, that segment is then considered as a neighbor of the current segment being analyzed. Once the neighbors are found, the second step is to compute the relative width and height of a segment with regard to the width and height of its neighbors. To do this, each segment's width and height are divided by the maximum width and height found in the neighborhood. Finally, for each segment, the \textit{Spatial Analysis} step outputs a \texttt{\small JSON} object containing the information regarding neighborhood coordinates, neighbor segments and relative size (width and height).

\subsection{Phase 4: DP Resolution}
\label{sec: DP Resolve}

The final phase of our approach is the \textit{DP Resolution} phase where our main goal is to identify the underlying DPs that are most likely to be present in the given UI. \textit{DP Resolution} is a two step process, consisting of \textit{Segment Level Resolution} and \textit{UI Level Resolution}. In \textit{Segment Level Resolution}, our approach identifies potential DPs in UI segments by considering the textual, color, and spatial analysis results (section \ref{sec: DP Analysis}). In the \textit{UI Level Resolution}, results from both \textit{Visual Cue Detection} (section \ref{sec: Visual Cue Detection}) and \textit{Segment Level Resolution} (section \ref{sec: Segment Level Resolve}) are used to to predict a final set of underlying DPs in the UI. Localization is performed using the bounding boxes of the identified UI elements from the screen.

\subsubsection{Segment Level Resolution}
\label{sec: Segment Level Resolve} To identify potential DPs in a segment, we take into consideration the results from textual, color, and spatial analyses (section \ref{sec: DP Analysis}). We employ a voting mechanism among the neighbor segments to resolve the most likely DPs at the segment level. In this process, a UI segment gets votes for a DP category from its neighbor segments if some specific textual, color, or spatial criteria are met. For instance, in case of text based resolution, a segment gets a vote from a neighbor segment if both of them exhibit textual patterns related to a similar DP. In the visual resolution, a segment gets a vote from its neighbor if the color brightness of the segment and the neighbor is opposite (\ie brighter/darker). In the spatial resolution, a segment gets a vote from a neighbor if the difference of the relative height or width between the segment and the neighbor is more than a predefined threshold value. In the voting process, we not only calculate the number of votes but also compute a score for the earned votes. The score of the votes depends on the number of factors or cues satisfied in textual, color, and spatial resolution processes. 

In summary, the voting process described above is akin to an equally weighted score across four features: (i) whether a target text pattern matched (\textit{text analysis}); (ii) whether text patterns in neighbor segments matched (\textit{text analysis}); (iii) whether there is high contrast in comparison to neighbor components (\textit{color analysis}); and (iv) whether there is a size difference in comparison to neighbor components (\textit{spatial analysis}). If the feature is present a 1 is assigned, and if a feature is not present, a 0 is assigned. The scores are calculated for the DP categories that correspond to the matched textual features from the analysis phase, for the individual features (textual/visual/color) defined for them. The scores are normalized depending upon how many analyses are relevant for a given DP. Next, in the UI level resolution, the scores are combined with the icon detection results using an 80/20 (scores/icon) ratio. The final output of the \textit{Segment Level Resolution} is a \texttt{\small JSON} object where for each segment a set of DPs with corresponding votes and scores are reported.

\subsubsection{UI Level Resolution} In this step, \AidUI makes the final prediction regarding any underlying DPs that are likely to be present in a given UI. To accomplish this, results from both \textit{Segment Level Resolution} (section \ref{sec: Segment Level Resolve}) and \textit{Visual Cue Detection} (section \ref{sec: Visual Cue Detection}) are taken as input. For each identified DP category, we take the votes with the highest score for given DP category found in \textit{Segment Level Resolution}. Later, if applicable, we also take into account the vote and confidence score from \textit{Visual Cue Detection}. In the case that both types of information are present, we combine the confidence scores from the \textit{Segment Level Resolution} and the \textit{Visual Cue Detection} using an 80/20 ratio (which we derived empirically). The output of \textit{UI Level Resolution} is a JSON object containing identified DP categories along with the number of votes, scores, and associated segment information for localization. Finally, the top (up to) two DP categories having the highest number of votes along with a certain confidence level are selected as the most likely DP.

\section{Evaluation Methodology}
\label{sec: Design of the Experiments}

In this section, we present the design of the study we employed for the evaluation of \AidUI. Our empirical study is aimed at assessing the \textit{performance} and \textit{robustness} of the approach as well as the contribution of different analysis modules to the overall effectiveness. To accomplish these study goals we formulated the following research questions:

\begin{itemize}
    \item \textbf{RQ$_1$:} \textit{What is the precision, recall and F1-score of \AidUI in detecting DP in UIs?}

    \item \textbf{RQ$_2$:} \textit{How robust is \AidUI in detecting DPs in UIs from different domains (mobile/web)?}

	\item \textbf{RQ$_3$:} \textit{How often does \AidUI detect false positives in screens that known to not contain DPs?}

    \item \textbf{RQ$_4$:} \textit{How well is \AidUI able to localize DPs in UI screens?}

    \item \textbf{RQ$_5$:} \textit{What is the contribution of different analysis modules in detecting DPs in UIs ?}

\end{itemize}

\subsection{Derivation of the \ContextDP Dataset}  
\label{sec: Dataset} 
To evaluate \AidUI, we have derived \ContextDP, the current largest dataset of labeled UIs containing both DP and Non-DP instances. \ContextDP includes total 501 mobile and web UI screenshots that represent 301 DP instances as well as 243 Non-DP instances.

To collect the mobile UIs, we made use of the comprehensive video dataset by Geronimo ~\etal~\cite{di2020ui} which includes mobile app usage screen recordings and classifications of identified DPs at various timestamps in the videos. The publicly available dataset by Geronimo~\etal~\cite{di2020ui} includes 15 videos of mobile apps and information about the timestamps of observed DPs in those videos.  Based on the available videos and the timestamps data, we extract the frames from the videos according to given timestamps. As the public dataset only contains 15 videos, we contacted and worked with the authors of the paper to extract all relevant DPs from their entire corpus of user videos, while carefully excluding video segments that may contain personal information. During this process, we collected 2994 UI screenshots in total spanning over 68 apps from 3 categories (communication, entertainment, and music). 

Similarly, based on the data provided by Mathur~\etal~\cite{mathur2019dark}, we collected over 1500 web screenshots from popular shopping websites with known DPs by leveraging their publicly available dataset. To further bolster the number of DP examples and the generalizability of our dataset, we also randomly collected a combined 5000 mobile UI screenshots from the RICO dataset ~\cite{Deka:2017:Rico} (not used to the train the icon detection model) and web UI screenshots from popular Alexa 100 shopping websites respectively. After the curation process, we randomly selected a total of 501 (339 mobile and 162 web) screenshots for labeling. This sampling represents a 95\% confidence level and 8\% confidence interval for mobile apps, and 12\% confidence interval for web UIs.

Next, three authors of this paper participated in a rigorous labeling process for both categorizing and creating bounding boxes for each observed DP in our sampled dataset. In order to ensure a consistent and agreed upon labeling strategy, prior to the labeling process, we conducted a comprehensive discussion among the authors to review the rationales behind the labeling decisions of our 501 UI screenshots. This process included examining and discussing two-three examples of each of the 10 targeted DPs from our taxonomy, and deriving a set of labeling and bounding box guidelines (these guidelines are available in our online appendix~\cite{appendix}). Then, two of the authors independently labeled each of the 501 UI screens using the Label Studio \cite{labelstudio} application, resolving necessary conflicts. The labeling procedure took place over three rounds, wherein the agreement of the last round was higher than 80\%, indicating strong agreement among the labelers.

\subsection{Evaluation Metrics}
\label{sec: Evaluation Metrics}
We evaluate \AidUI with respect to all RQs by using following metrics: \textit{precision}, \textit{recall}, \textit{F1-score}, and \textit{IoU}. The metrics are defined as follows:
\begin{itemize}
    \item \textbf{Precision} describes the ability of the classifier to properly distinguish between true positives and false positives. \textit{Precision} is computed as $\textit{Precision} = \textit{TP}/\textit{TP + FP} $, where TP is the number of true positives and FP is the number of false positives. In our context a TP (true positive) is a correct prediction that a given DP type exists on a given UI screen and matches the ground truth, and an FP (false positive) is when the approach predicts a DP type that is not present on a given screen in \ContextDP.
    \item \textbf{Recall} is intuitively the ability of the classifier to find all the samples that are positive. \textit{Recall} is computed as $\textit{Recall} = \textit{TP}/\textit{TP + FN}$, where TP is the number of true positives and FN is the number of false negatives. In this context a false negative is a case when a DP is present on a screen, but \AidUI could not detect it.
    \item \textbf{F1-score} is the harmonic mean of the precision and recall. \textit{F1-score} is defined as $\textit{F1-score} =\textit{2 $\times$ Precision $\times$ Recall}/\textit{Precision + Recall}$
    \item \textbf{IoU} computes the amount of overlap between the predicted and ground truth bounding boxes. \textit{IoU} is defined as: $\textit{IoU} = \textit{Area of Overlap}/\textit{Area of Union}$. Note that we define two versions of \textit{IoU} for our evaluation. First, \textit{strict IoU} operates according to the formula above, wherein predictions are directly compared to the ground truth. Second we define a \textit{contained IoU} wherein we set the \textit{IoU} to 100\% if the predicted bounding box falls \textit{within} the ground truth bounding box. This is because during our investigation of the localization results, we found that \AidUI often predicts more specific bounding boxes for a given DP that could still be helpful to end users. Thus, this measure may offer a more realistic characterization of \AidUIs performance by rewarding segments identified within the ground truth bounding box.
\end{itemize}

\subsection{RQ$_1$ \& RQ$_2$: DP Detection Performance Across Domains}

To answer RQ$_1$, we  assess the ability of \AidUI to accurately detect DP instances. This experiment aims to measure \AidUI's performance both as a whole, and for individual DP categories in order to determine \AidUI's effectiveness on individual DP categories. In this process, we evaluate \AidUI on \ContextDP (section \ref{sec: Dataset}) using the aforementioned evaluation metrics (\textit{precision, recall and F1-score}) described in section \ref{sec: Evaluation Metrics}. The evaluation metrics are computed using the predictions and derived ground truth from \ContextDP. To answer RQ$_2$ we compare the detection results across both the web and mobile portions of the \ContextDP dataset.

\begin{table*}[]
\small
\vspace{-1em}
\caption{\AidUI DP Category-wise Detection/Classification Performance}
\label{tab:category-wise-performance-classification-all-mobile-web}
\resizebox{\textwidth}{!}{
\begin{tabular}{@{}lllllllllllllllllllllll@{}}

\toprule
\multicolumn{1}{c}{} & \multicolumn{4}{c}{\textbf{All}} &      & \multicolumn{4}{c}{\textbf{Mobile}} &      & \multicolumn{4}{c}{\textbf{Web}} \\ 

\midrule
\textbf{DP Category} & \textbf{Instances} & \textbf{Precision} & \textbf{Recall} & \textbf{F1-score} &      & \textbf{Instances} & \textbf{Precision} & \textbf{Recall} & \textbf{F1-score} &      & \textbf{Instances} & \textbf{Precision} & \textbf{Recall} & \textbf{F1-score}\\
Activity Message & 10 & 1.0 & 0.80 & 0.89 &      & -- & -- & -- & -- &      & 10 & 1.0 & 0.80 & 0.89\\
High Demand Message & 5 & 1.0 & 0.80 & 0.89 &      & -- & -- & -- & -- &      & 5 & 1.0 & 0.80 & 0.89\\
Low Stock Message & 19 & 0.75 & 0.79 & 0.77 &      & -- & -- & -- & -- &      & 19 & 0.75 & 0.79 & 0.77\\
Limited Time Message & 26 & 0.95 & 0.73 & 0.83 &      & -- & -- & -- & -- &      & 26 & 0.95 & 0.73 & 0.83\\
Countdown Timer & 28 & 0.63 & 0.86 & 0.73 &      & -- & -- & -- & -- &      & 28 & 0.83 & 0.86 & 0.84\\
Attention Distraction & 13 & 0.55 & 0.46 & 0.50 &      & 9 & 0.29 &  0.22 & 0.25 &      & 4 & 1.0 & 1.0 & 1.0\\
Default Choice & 111 & 0.67 & 0.59 & 0.62 &      & 99 & 0.73 & 0.60 & 0.66 &      & 12 & 0.38 & 0.50 & 0.43\\
Disguised Ads & 21 & 0.32 & 0.57 & 0.41 &      & 21 & 0.33 & 0.57 & 0.42 &      & -- & -- & -- & --\\
Nagging & 57 & 0.52 & 0.77 & 0.62 &      & 57 & 0.54 & 0.77 & 0.64 &      & -- & -- & -- & --\\
Gamification & 11 & 0.80 & 0.36 & 0.50 &      & 11 & 1.0 & 0.36 & 0.53 &      & -- & -- & -- & --\\

\midrule
\textbf{Total Instances} & \multicolumn{4}{c}{\textbf{301}} &  & \multicolumn{4}{c}{\textbf{197}} &  & \multicolumn{4}{c}{\textbf{104}} \\
\textbf{Avg. Precision} & \multicolumn{4}{c}{\textbf{0.66}} &  & \multicolumn{4}{c}{\textbf{0.63}} &  & \multicolumn{4}{c}{\textbf{0.82}} \\
\textbf{Avg. Recall} & \multicolumn{4}{c}{\textbf{0.67}} &  & \multicolumn{4}{c}{\textbf{0.61}} &  & \multicolumn{4}{c}{\textbf{0.77}} \\ 
\textbf{Avg. F1-score} & \multicolumn{4}{c}{\textbf{0.65}} &  & \multicolumn{4}{c}{\textbf{0.60}} &  & \multicolumn{4}{c}{\textbf{0.79}} \\

\bottomrule
\end{tabular}
}
\vspace{.5em}
\vspace{-1em}

\end{table*}

\subsection{RQ$_3$: Potential False Positive DP Detections}

\AidUIs utility as a potential developer tool is predicated on the intention that it perform reasonably well at both detecting dark patterns when they do exist, and not triggering false alarms for screens that do not contain DP instances. Thus, to evaluate the potential of \AidUI to trigger false positives, we applied our technique to the 243 screenshots that do not contain DPs from \ContextDP.

\subsection{RQ$_4$: Localization Performance}
We answer RQ$_4$ using both \textit{contained} and \textit{strict} IoU which are calculated by measuring the difference between the predicted bounding boxes from \AidUI and the ground truth bounding boxes from \ContextDP.

\subsection{RQ$_5$: Ablation Study of DP Analyses}
\label{sec: Ablation Study of DP Analyses}
To answer this RQ, we conduct an experiment that is aimed at exploring the contributions of the textual, color, and spatial analysis modules in detecting DPs. To accomplish this, we choose different combinations of modules while conducting the evaluation process. As text analysis acts as the base module, it is selected in all the combinations. The combinations that we use in the evaluation are: \textit{Text}, \textit{Text + Color Analysis}, \textit{Text + Spatial Analysis}, and \textit{Text + Color + Spatial Analysis}. For each combination, we calculate the same evaluation metrics that we use in RQ1.

\section{Empirical Results}
\label{sec: Results}

\begin{table}[]
\small
\caption {\AidUI Overall Detection/Classification Performance}
\label{tab:overall-performance-classification-dp-nodp}
\begin{tabular}{@{}lllll@{}}
\toprule
\textbf{} & \textbf{Instances} & \textbf{Precision} & \textbf{Recall} & \textbf{F1-score} \\
\textbf{Non-DP} & 243 & 0.85 & 0.86 & 0.85 \\
\textbf{DP} & 301 & 0.66 & 0.67 & 0.65 \\
\bottomrule
\end{tabular}
\centering
\vspace{.5em}
\end{table}

\subsection{RQ$_1$: Detection Performance}
Our main goal in RQ1 is to measure the performance of \AidUI in terms of detection/classification of both Non-DP and DP instances. In answering this RQ, we are interested in assessing \AidUIs overall performance as well as performance specific to individual DP categories. We conduct the performance evaluation based on the metrics stated in section \ref{sec: Evaluation Metrics}, \ie \textit{precision}, \textit{recall} and \textit{F1-score}. Table \ref{tab:overall-performance-classification-dp-nodp} and \ref{tab:category-wise-performance-classification-all-mobile-web} illustrate the aggregate and category-wise classification performance respectively.

\begin{table*}[]
\vspace{-1.5em}
\caption{\AidUI Localization Performance}
\label{tab:performance-localization-all-mobile-web}
\small
\resizebox{\textwidth}{!}{
\begin{tabular}{@{}lllllllllll@{}}
\toprule
\multicolumn{1}{c}{} 
& \multicolumn{2}{c}{\textbf{All}} &  &  
& \multicolumn{2}{c}{\textbf{Mobile}} &  &  
& \multicolumn{2}{c}{\textbf{Web}} \\ 

\midrule
\textbf{DP Category} 
& \textbf{Avg. Strict IoU} 
& \textbf{Avg. Contained IoU} &  &  
& \textbf{Avg. Strict IoU} 
& \textbf{Avg. Contained IoU} &  &  
& \textbf{Avg. Strict IoU} 
& \textbf{Avg. Contained IoU} \\

Activity Message & 0.351397 & 0.740450 &  &  & -- & -- &  &  & 0.351397 & 0.740450 \\
High Demand Message & 0.340726 & 0.730416 &  &  & -- & -- &  &  & 0.340726 & 0.730416 \\
Low Stock Message & 0.223910 & 1.000000 &  &  & -- & -- &  &  & 0.223910 & 1.000000 \\
Limited Time Message & 0.262617 & 0.808963 &  &  & -- & -- &  &  & 0.262617 & 0.808963 \\
Countdown Timer & 0.231990 & 0.823345 &  &  & -- & -- &  &  & 0.231990 & 0.823345 \\

Attention Distraction & 0.040522 & 1.000000 &  &  & 0.081790 & 1.000000 &  &  & 0.019887 & 1.000000 \\
Default Choice & 0.023972 & 0.569231 &  &  & 0.024030 & 0.542373 &  &  & 0.023409 & 0.833333 \\
Disguised Ads & 0.006635 & 0.916667 &  &  & 0.006635 & 0.916667 &  &  & -- & -- \\
Nagging & 0.015434 & 0.957007 &  &  & 0.015434 & 0.957007 &  &  & -- & -- \\
Gamification & 0.125986 & 0.777633 &  &  & 0.125986 & 0.777633 &  &  & -- & -- \\

\midrule
\textbf{Overall} 
& \textbf{0.162319} 
& \textbf{0.832371} &  &  
& \textbf{0.050775} 
& \textbf{0.838736} &  &  
& \textbf{0.207705} 
& \textbf{0.848072} \\ 

\bottomrule
\end{tabular}
}
\vspace{.5em}
\vspace{-1em}

\end{table*} 

From the classification results in table \ref{tab:overall-performance-classification-dp-nodp}, we observe that \AidUI achieves an overall \textit{average precision} of 66\%, \textit{average recall} of 67\% and \textit{average F1-score} of 65\% in detecting DP instances.
Moreover, category-wise results in Table \ref{tab:category-wise-performance-classification-all-mobile-web} show that a subset of DPs (\eg \texttt{\small Activity Message}, \texttt{\small High Demand Message}, \texttt{Low Stock Message}, \texttt{\small Limited Time Message}, \texttt{\small Countdown Timer} etc.) can reliably be detected with moderate to high \textit{precision}, \textit{recall} and \textit{F1-score} values.

\begin{figure}
\centering
\vspace{-1em}
\includegraphics[width=0.8\linewidth]{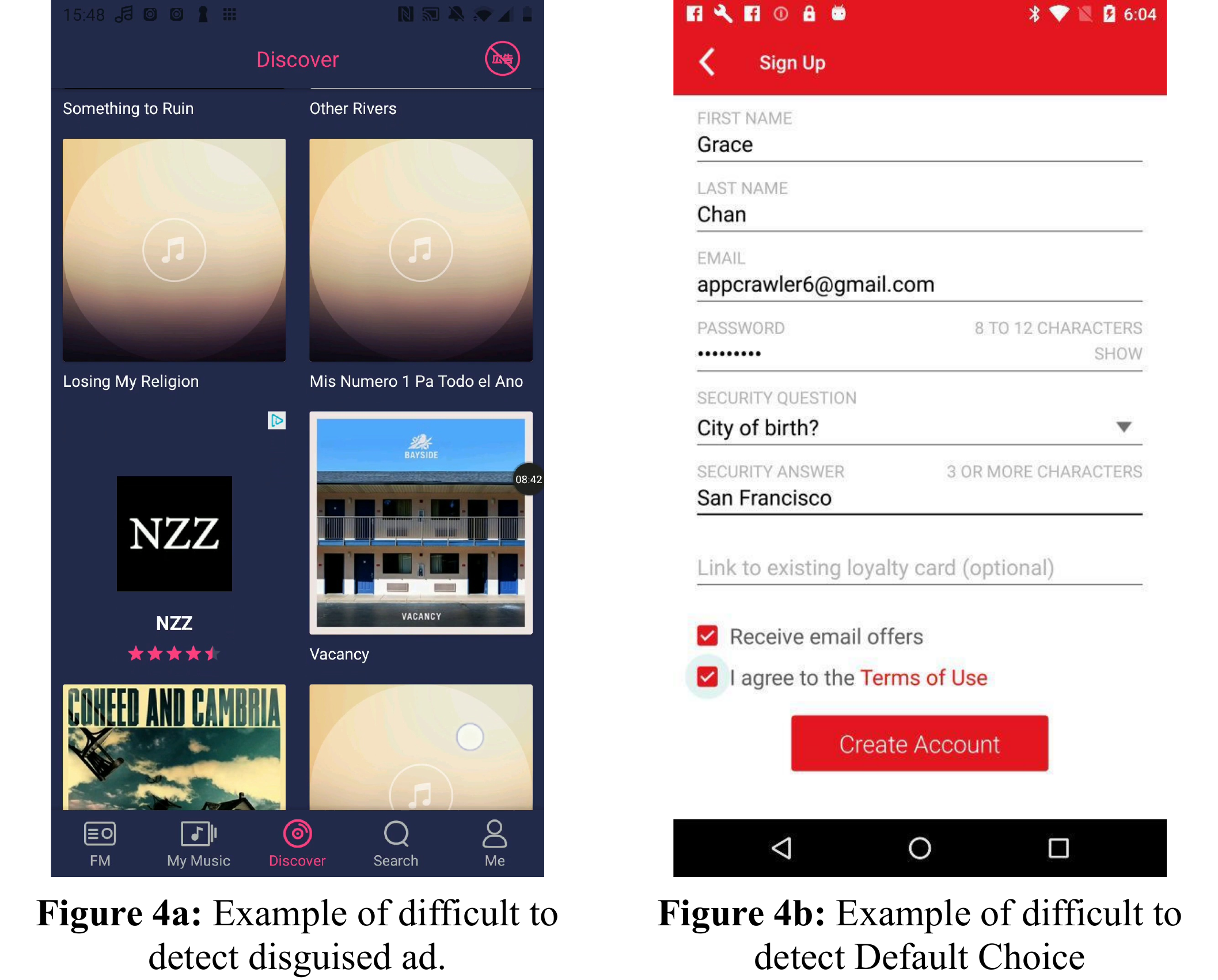}
\caption{Example of difficult to detect Dark Patterns}
\label{fig:difficult_dark_patterns}
\end{figure}

Instances where \AidUI fails to identify DPs are mostly due to deviations in textual or visual patterns exhibited by some of the DP categories. For instance, 32\% of the \texttt{\small Disguised Ads} instances are correctly predicted whereas 26\% of them are wrongly predicted as \textit{Non-DP}. To better understand, we present one such representative instance of \textit{Disguised Ads} in Figure \ref{fig:difficult_dark_patterns}. Here, at a first glance, the appearance of the advertisement looks very similar to regular content. Though on the top right corner there is an \textit{ad} icon, this particular advertisement has a subtle difference compared to other typical advertisements. To detect such subtle visual cues, we may need to develop separate deep learning solution based on a comprehensive study of diverse types of UI advertisement contents. Similarly, for some DP categories our approach fails due to the limitation of the vocabulary or textual patterns that AidUI considers, as illustrated in Figure 4b for the \texttt{\small Default Choice} DP. Future work could focus toward resolving these limitations through the use of \textit{language modeling} wherein embeddings from the models are used to measure textual similarity, and hence would be able to better account for semantically similar, yet lexically varied text.

\vspace{-1em}
\begin{tcolorbox}[enhanced,skin=enhancedmiddle,borderline={1mm}{0mm}{MidnightBlue}]\small
\textbf{Answer to RQ$_1$}: \AidUI achieves an overall average \textit{precision}, \textit{recall} and \textit{F1-score} of 0.66, 0.67 and 0.65 respectively. However, for a subset of DP classes, including \texttt{\small Activity Message}, \texttt{\small High Demand Message}, \texttt{\small Low Stock Message}, \texttt{\small Limited Time Message} and \texttt{\small Countdown Timer}, \AidUI performs well with high average \textit{precision} ($\approx$0.87), \textit{recall} ($\approx$0.80) and \textit{F1-score} ($\approx$0.82).
\end{tcolorbox}

\begin{table*}[th]
\vspace{-1em}
\caption{Contribution of different modules}
\label{tab:contribution-modules}
\small
\resizebox{0.90\textwidth}{!}{
\begin{tabular}{@{}llllllllllll@{}}

\toprule
\multicolumn{1}{c}{} & \multicolumn{3}{c}{\textbf{All}} &  & \multicolumn{3}{c}{\textbf{Mobile}} &  & \multicolumn{3}{c}{\textbf{Web}} \\

\midrule
\multicolumn{1}{c}{} & \multicolumn{3}{c}{DP Instances: 301} &  & \multicolumn{3}{c}{DP Instances: 197} &  & \multicolumn{3}{c}{DP Instances: 104} \\

\midrule
\textbf{Modules} & \multicolumn{1}{c}{\textbf{\begin{tabular}[c]{@{}c@{}}Avg. \\ Precision\end{tabular}}} 
& \multicolumn{1}{c}{\textbf{\begin{tabular}[c]{@{}c@{}}Avg. \\ Recall\end{tabular}}} 
& \multicolumn{1}{c}{\textbf{\begin{tabular}[c]{@{}c@{}}Avg. \\ F1-score\end{tabular}}} &  & 
\multicolumn{1}{c}{\textbf{\begin{tabular}[c]{@{}c@{}}Avg. \\ Precision\end{tabular}}} 
& \multicolumn{1}{c}{\textbf{\begin{tabular}[c]{@{}c@{}}Avg. \\ Recall\end{tabular}}}
& \multicolumn{1}{c}{\textbf{\begin{tabular}[c]{@{}c@{}}Avg. \\ F1-score\end{tabular}}} &  & 
\multicolumn{1}{c}{\textbf{\begin{tabular}[c]{@{}c@{}}Avg. \\ Precision\end{tabular}}} 
& \multicolumn{1}{c}{\textbf{\begin{tabular}[c]{@{}c@{}}Avg. \\ Recall\end{tabular}}}
& \multicolumn{1}{c}{\textbf{\begin{tabular}[c]{@{}c@{}}Avg. \\ F1-score\end{tabular}}}\\

Text  & 0.75 & 0.44 & 0.40 &  & 0.75 & 0.32 & 0.27 &  & 0.74 & 0.67 & 0.70 \\
Text + Color  & 0.78 & 0.45 & 0.42 &  & 0.78 & 0.33 & 0.29 &  & 0.74 & 0.67 & 0.70 \\
Text + Spatial  & 0.65 & 0.65 & 0.63 &  & 0.63 & 0.61 & 0.60 &  & 0.78 & 0.73 & 0.75 \\
Text + Color + Spatial & 0.66 & 0.67 & 0.65 &  & 0.63 & 0.61 & 0.60 &  & 0.82 & 0.77 & 0.79 \\ 

\midrule
\end{tabular}
}
\centering
\vspace{.5em}
\vspace{-2em}

\end{table*}

\subsection{RQ$_2$: Domain-specific Performance}

Table \ref{tab:category-wise-performance-classification-all-mobile-web} illustrates that \AidUI performs significantly better on web UIs as compared to the UIs from mobile applications. In fact, we observe an increase of 30\% for \textit{avg. precision}, 26\% for \textit{avg. recall} and 31\% for \textit{avg. F1-score} between the two portions of our dataset. Our insight regarding the significant performance difference in identifying DPs across mobile and web domain is related to the prevalence of different types of DPs in those domains. While developing our unified taxonomy, we observed that there are particular groups of DPs that have dominant presence in a particular domain, \ie web or mobile. Moreover, we also observed that the patterns frequently found in web UIs typically skew toward textual cues whereas several patterns that are dominant in mobile UIs often contain \textit{both} textual and visual cues. This phenomenon makes the DP identification task in mobile UIs more complex as compared to web UIs. As our current approach equally considers the outputs from text and color analysis, this could be one reason for the observed performance gap across the two software domains. Future work should aim to empirically examine the textual and visual differences in DPs across software domains in an effort to better inform future automated techniques.

\begin{tcolorbox}[enhanced,skin=enhancedmiddle,borderline={1mm}{0mm}{MidnightBlue}]\small
\textbf{Answer to RQ$_2$}: \AidUI shows significant difference in performance across web and mobile domains. This is likely due to the fact that the observed DPs in our mobile dataset contain more varied patterns that are more difficult to detect.
\end{tcolorbox}

\subsection{RQ$_3$: False Positive DP Detection}

To measure the rate of false positives, we applied \AidUI to a set of screens that, during the labeling process for ContextDP, were confirmed to not exhibit any dark patterns. Of the 243 to which we applied \AidUI, only 36 screens ($\approx$15\%) were identified as having false positives, and a vast majority of these fell into the \texttt{\small Default Choice} DP category. It should be noted that given the scoring procedure among the various components of \AidUI's approach, it is possible to calculate a confidence threshold that a given screen contains a DP, which could be used to help indicate the potential severity or confidence of predictions for future developer tools.

\begin{tcolorbox}[enhanced,skin=enhancedmiddle,borderline={1mm}{0mm}{MidnightBlue}]\small
\textbf{Answer to RQ$_3$}: \AidUI exhibits a false positive rate of 36/243 (14\%) when applied to screens confirmed to not exhibit DPs. However, most of these misclassifications are into a single class \texttt{\small Default Choice} DP class, and could be further mitigated both by adjusting the sensitivity of cues for this class, and by displaying confidence scores for given predictions for future developer-facing tools.
\end{tcolorbox}

\vspace{-1em}
\subsection{RQ$_4$: Localization Performance}

The results in Table \ref{tab:performance-localization-all-mobile-web} illustrate that our approach achieves a fairly low \textit{strict IoU} both overall and on a category-by-category basis. However, this measurement approach presents a skewed view of the practical performance of \AidUI. This is because after examining both predicted and ground truth bounding boxes we noticed that this is largely due to precise bounding box selection by our text analysis module. Though our approach is actually able to localize a precise bounding box for various UIs, it is getting penalized because of having smaller intersections as compared to the ground truth, which tended to encompass areas between text, for example. Based on our observation, we defined the \textit{contained IoU} to provide a more complete picture of \AidUIs localization performance. The overall \textit{avg. contained IoU} value 0.83 illustrates that \AidUI is properly localizing elements that exist within the ground truth bounding boxes, although the lablers of the \ContextDP dataset often felt that a \textit{larger} portion of the screen should be considered to contain given DPs, usually due to negative space between text.

\begin{tcolorbox}[enhanced,skin=enhancedmiddle,borderline={1mm}{0mm}{MidnightBlue}]\small
\textbf{Answer to RQ$_4$}: When examining \textit{strict IoU}, \AidUI performs poorly as it tends to localize smaller areas of the screen. However, when considering \textit{contained IoU}, we find that the areas of the screen that \AidUI localizes consistently fall within the ground truth for the DPs in \ContextDP.
\end{tcolorbox}

\subsection{RQ$_5$: Ablation Study of DP Analyses}
Finally, we conduct an ablation study to examine the contribution of \AidUIs textual, color and spatial analysis modules in detecting DPs. In this experiment, we remove one module at a time to understand the contribution of other modules. As stated earlier in section \ref{sec: Ablation Study of DP Analyses}, \textit{text analysis} serves as the foundation of our implemented approach. Hence, we include the \textit{text analysis} module in every combination of modules we use in our ablation study. We first start with using all three modules (\ie \textit{text + color + spatial}). In later steps, we removed \textit{color} and \textit{spatial} modules respectively. Finally, we end up with using \textit{text analysis} only. 

\begin{tcolorbox}[enhanced,skin=enhancedmiddle,borderline={1mm}{0mm}{MidnightBlue}]\small
\textbf{Answer to RQ$_5$}: \AidUI \revision{achieves best results} when all three, \ie \textit{text}, \textit{color} and \textit{spatial}, analysis modules are present. This suggests that there are orthogonal features that exist across data modalities that are useful for automated DP detection.
\end{tcolorbox}

\section{Related Work}
\label{sec:related-work}

\subsection{Automated Detection of Dark Patterns}
Raju~\etal~\cite{10.1007/978-981-16-5157-1_72} proposed an intended design aimed to analyze the source code of a loaded web page to detect the advertisements that correspond to certain dark pattern types. Another work by Liu~\etal~\cite{10.1145/3366423.3380242} proposed a framework that leverages automated app testing with ad traffic identification approach to detect devious ad contents. Unlike these works, which are solely focused on detecting suspicious advertisements, our approach is aimed to detect a wide range of different types of dark patterns coming from both mobile and web applications.

\subsection{Automated UI Understanding \& Detection of Design Issues}

Wu \etal introduced Screen Parsing~\cite{Wu:UIST'21}, and Chen \etal introduced UIED~\cite{chen2020object}, both of which use  computer vision and deep learning tehcniques to segment and classify UI elements from screenshot pixels. Zhang \etal and Chen \etal developed approaches for screen content recognition~\cite{Zhang:CHI'21} and icon labeling~\cite{Chen:ICSE'20} that can automatically infer accessibility metadata from screen pixels. Moran~\etal introduced {\sc ReDraw}~\cite{Moran:TSE'18}, which uses a combination of unsupervised computer vision and deep learning techniques to automatically prototype UI code for mobile apps from mock-ups.
\AidUI differs from these above techniques in two major ways. First, the screen properties which are identified by \AidUI (DPs) are novel compared to properties inferred by past work (\eg accessibility data, screen elements). Second, due to the nature of these properties, \AidUI employs different analysis techniques. While a majority of the techniques above used end-to-end deep learning, this is difficult given that dark pattern examples need to manually sourced, making large scale data collection challenging. Thus, \AidUI processes visual and textual cues, as opposed to training a purely deep learning-based classification solution -- making it largely \textit{complementary} to existing work. 

Additionally, the software engineering research community has been working towards identifying UI display issues across multiple types of software including web~\cite{Mahajan:ICST'18,Mahajan:ICST'16}, and mobile apps~\cite{Yang:ICSE'21,Liu:ASE'20,Zhao:ICSE'20,Moran:ICSE'18}. These techniques tend to use a combination of deep learning and unsupervised computer vision techniques to detect varying types of display issues such as design guideline violations, internationalization issues, and deviations from design specifications. However, none of these techniques is capable of detecting dark patterns.

\subsection{Ethics and Dark Patterns in UI/UX Design}
A longstanding goal of the broader HCI research community has been to develop effective frameworks and guidelines to improve the UI/UX of applications
Hence, investigating the ethical aspects of UI/UX from designer's perspectives is a growing area of interest in the HCI community and researchers have already explored a number of framings regarding ethics and values
\cite{bardzell2013critical, flanagan2014values, fogg2009behavior, frauenberger2017action, friedman2003human, sengers2005reflective, shilton2013values, shilton2014see}. In this paper, we aim to build upon past research done in the topics of DPs and ethical UI/UX design by investigating how automated approaches for DP detection may equip developers with the tools necessary to understand/avoid deceptive designs.

\section{Limitations \& Threats to Validity}
\label{sec:threats}

% \noindent \textit{\textbf{Internal \& Construct Validity:}} One potential threat to validity is incorrect labels or bounding boxes in ContextDP. However, we mitigate this threat by following a rigorous labeling methodology with multiple author agreement. An additional threat is related to the completeness of \AidUIs textual and visual cues. However, these cues were derived empirically through examining a small set of DP examples, and our experiments illustrate that they function reasonably well.

% \noindent \textit{\textbf{External Validity:}} We developed the current largest and most diverse set of fully localized DPs across two application domains (web + mobile). Despite this, there is a potential threat to validity that \AidUI may not generalize beyond the \ContextDP dataset.

\noindent \textit{\textbf{Limitations:}} \AidUI has various limitations that serve as motivation for future research. First, our the current implementation of \AidUI is targeted toward 10 DP categories, which means future work is required to developed automated techniques to detect other categories. As already discussed in section~\ref{background}, our current approach is focused on working with DP categories that are manifested by different visual and textual cues on a single screen. Thus, one clear future research direction is to develop detection mechanisms for dynamic DP categories that are characterized by multiple screens and user actions.

Another limitation, which we plan to address in future work, is that our current text analysis technique is solely based on heuristically defined pattern-matching rules and hence are difficult to apply to certain DP categories (\eg \texttt{\small toying with emotion}) that involve semantically complex textual patterns to persuade the user into a particular action. As already discussed in section \ref{sec: Results}, the current text analysis approach has also the limitation of not detecting semantically similar, yet lexically varied text. We plan to leverage neural language models  to address these challenges in future work.

\noindent \textit{\textbf{Internal Validity:}} Threats to internal validity correspond to unexpected factors in the experiments that may contribute to observed results. The main threat to internal validity is related to the construction of the ContextDP dataset. However, we mitigated this threat by creating the dataset through an independent multi-author labeling procedure with high agreement. Another potential threat to internal validity is the construction of our training dataset for the FasterRCNN \cite{ren2015faster}. However, we followed best practices of prior work~\cite{cardenas-translating} and our trained model achieves high accuracy, mitigating this threat.

\noindent \textit{\textbf{Construct Validity:}} Threats to construct validity concern the operationalization of experimental artifacts. The main threat to construct validity is related to the measuring of the results of our approach in comparison to the ground truth. However, given that our approach outputs bounding boxes and DP labels, we were able to directly and automatically compute results.

\noindent \textit{\textbf{External Validity:}} Threats to external validity concern the generalization of the results. The main threats to our approach relate to the generalizability of the heuristics used to detect dark patterns. However, we derived these heuristics using 2-3 examples of each DP not included in ContextDP, and they appear to generalize well to the larger ContextDP dataset. Another threat to external validity is related to the generalizability of the ContextDP dataset itself. While we do not claim that our results generalize beyond this dataset, we did mitigate this threat by including a large number of examples across two different software domains (web and mobile).

\section{Conclusion}
\label{sec:conclusion}

In this paper, we have taken the first steps toward investigating the feasibility of automated detection of UI dark patterns in mobile and web UIs. We unified similar DP categories from existing taxonomies together and derived \ContextDP, the largest current fully labeled and localized DP instances. Furthermore, we implemented \AidUI, a fully-automated DP detection and localization technique, which performs well on ContextDP. Our results show that \AidUI performs well, in terms of \textit{precision}, \textit{recall} and \textit{F1-score} on a large subset of our studied DPs. We also illustrated that automated DP detection techniques appear to benefit from fusing multimodal data (\eg text and visual information). The summative findings of this work illustrate the feasibility and promise of automated DP detection/localization techniques. %that could serve as a light in the darkness for millions of users who must regularly grapple with deceptive UI designs.

\section*{acknowledgements}

%Acks
This work is supported in part by the NSF grants: CCF-2132285 and CCF-1955853. Any opinions, findings, and conclusions expressed herein are the authors and do not necessarily reflect those of the sponsors.

\IEEEpeerreviewmaketitle

\balance
\pagebreak
\bibliographystyle{abbrv}
\bibliography{references.bib}

\end{document}